\documentclass[manuscript,screen]{acmart}

\AtBeginDocument{%
  }

\setcopyright{acmcopyright}
\copyrightyear{2022}
\acmYear{2022}
\acmDOI{XXXXXXX.XXXXXXX}

\acmPrice{15.00}
\acmISBN{978-1-4503-XXXX-X/18/06}

\usepackage{bm}
\usepackage{subcaption}

\usepackage{multirow}
\usepackage{mathtools,amssymb,euscript,yfonts,latexsym}
\usepackage[capitalise]{cleveref}
\usepackage{makecell}

\usepackage{color,colortbl}
\usepackage{textcomp}

\definecolor{LightCyan}{rgb}{0.8,1,1}
\definecolor{Gray}{gray}{0.9}

\renewcommand{\paragraph}[1]{\vspace{1.25mm}\noindent\textbf{#1}}

\makeatletter
\usepackage{xspace}
\def\@onedot{\ifx\@let@token.\else.\null\fi\xspace}
\DeclareRobustCommand\onedot{\futurelet\@let@token\@onedot}

\begin{document}

\title{One Small Step for Generative AI, One Giant Leap for AGI: A Complete Survey on ChatGPT in AIGC Era}

\author{Chaoning Zhang}
\affiliation{%
  \institution{Kyung Hee University}
  \country{South Korea}
}
\email{chaoningzhang1990@gmail.com}

\author{Chenshuang Zhang}
\affiliation{%
  \institution{KAIST}
  \country{South Korea}
}
\email{zcs15@kaist.ac.kr}

\author{Chenghao Li}
\affiliation{%
  \institution{KAIST}
  \country{South Korea}
}
\email{lch17692405449@gmail.com}

\author{Yu Qiao}
\affiliation{%
  \institution{Kyung Hee University}
  \country{South Korea}
}
\email{qiaoyu@khu.ac.kr}

\author{Sheng Zheng}
\affiliation{%
  \institution{Beijing Institute of Technology}
  \country{China}
}
\email{zszhx2021@gmail.com}

\author{Sumit Kumar Dam}
\affiliation{%
  \institution{Kyung Hee University}
  \country{South Korea}
}
\email{skd160205@khu.ac.kr}

\author{Mengchun Zhang}
\affiliation{%
  \institution{KAIST}
  \country{South Korea}
}
\email{zhangmengchun527@gmail.com}

\author{Jung Uk Kim}
\affiliation{%
  \institution{Kyung Hee University}
  \country{South Korea}
}
\email{ju.kim@khu.ac.kr}

\author{Seong Tae Kim}
\affiliation{%
  \institution{Kyung Hee University}
  \country{South Korea}
}
\email{st.kim@khu.ac.kr}

\author{Jinwoo Choi}
\affiliation{%
  \institution{Kyung Hee University}
  \country{South Korea}
}
\email{jinwoochoi@khu.ac.kr}

\author{Gyeong-Moon Park}
\affiliation{%
  \institution{Kyung Hee University}
  \country{South Korea}
}
\email{gmpark@khu.ac.kr}

\author{Sung-Ho Bae}
\affiliation{%
  \institution{Kyung Hee University}
  \country{South Korea}
}
\email{shbae@khu.ac.kr}

\author{Lik-Hang Lee}
\affiliation{%
  \institution{Hong Kong Polytechnic University}
  \country{Hong Kong SAR (China)}
}
\email{lik-hang.lee@polyu.edu.hk}

\author{Pan Hui}
\affiliation{%
  \institution{Hong Kong University of Science and Technology (Guangzhou)}
  \country{China}
}
\email{panhui@ust.hk}

\author{In So Kweon}
\affiliation{%
  \institution{KAIST}
  \country{South Korea}
}
\email{iskweon77@kaist.ac.kr}

\author{Choong Seon Hong}
\affiliation{%
  \institution{Kyung Hee University}
  \country{South Korea}
}
\email{cshong@khu.ac.kr}

\renewcommand{\shortauthors}{Zhang et al.}


\begin{abstract}

OpenAI has recently released GPT-4 (a.k.a. ChatGPT plus), which is demonstrated to be one small step for generative AI (GAI), but one giant leap for artificial general intelligence (AGI). Since its official release in November 2022, ChatGPT has quickly attracted numerous users with extensive media coverage. Such unprecedented attention has also motivated numerous researchers to investigate ChatGPT from various aspects. According to Google scholar, there are more than 500 articles with ChatGPT in their titles or mentioning it in their abstracts. Considering this, a review is urgently needed, and our work fills this gap. Overall, this work is the first to survey ChatGPT with a comprehensive review of its underlying technology, applications, and challenges. Moreover, we present an outlook on how ChatGPT might evolve to realize general-purpose AIGC (a.k.a. AI-generated content), which will be a significant milestone for the development of AGI. 

\end{abstract}





\maketitle
{
  \hypersetup{linkcolor=black}
  \tableofcontents
}

\section{Introduction}\label{sec:introduction}

\begin{figure*}
    \centering
    \includegraphics[width=\linewidth]{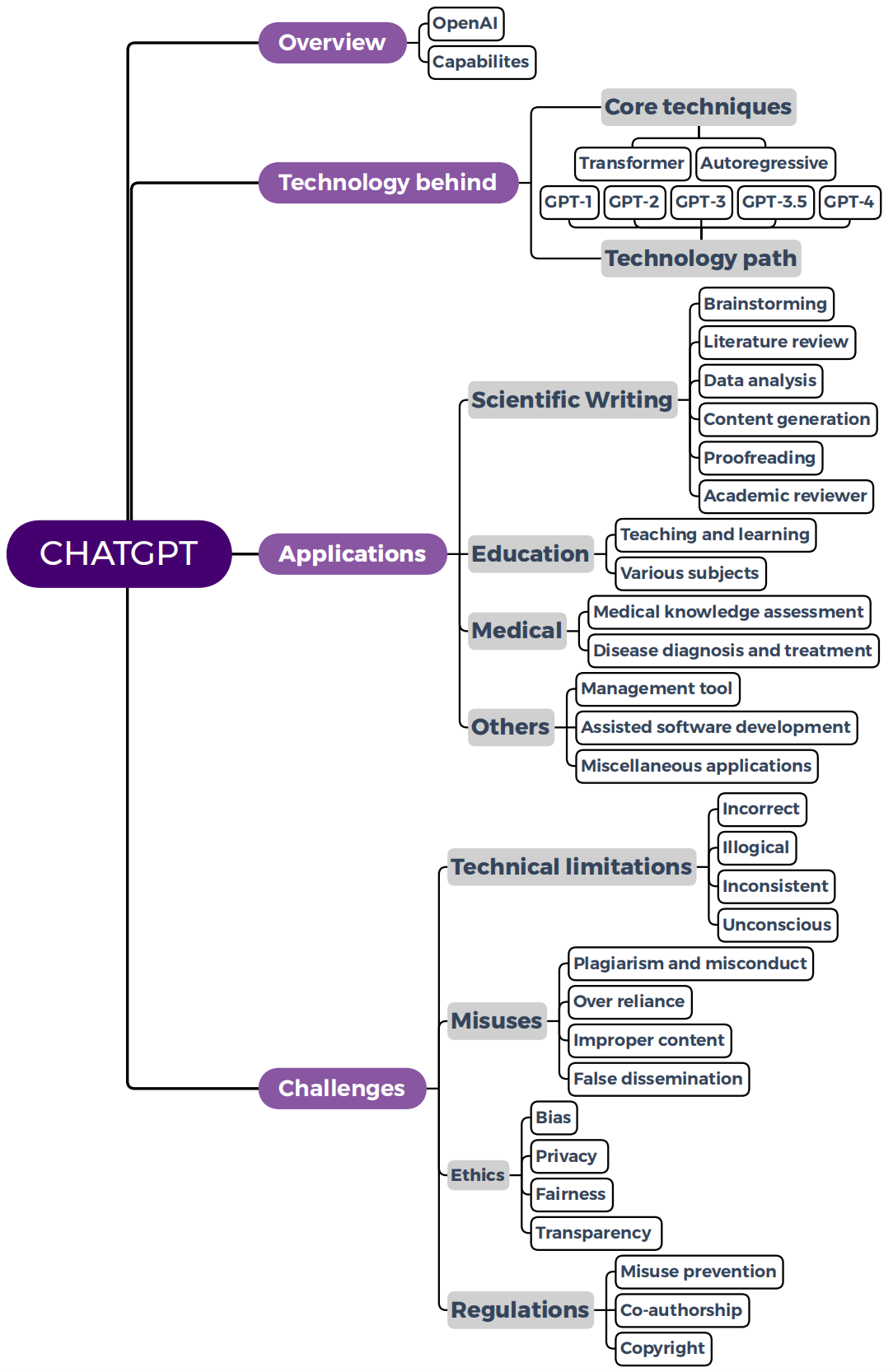}
    \caption{Structure overview of this survey.}
    \label{fig:Overview}
\end{figure*}

The past few years have witnessed the advent of numerous generative AI (AIGC, a.k.a. AI-generated content) tools~\cite{ramesh2021zero,radford2022robust,ji2020comprehensive}, suggesting AI has entered a new era of creating instead of purely understanding content. For a complete survey on generative AI (AIGC), the readers can refer to~\cite{zhang2023complete}. Among those AIGC tools, ChatGPT, which was released in November 2022, has caught unprecedented attention. It attracted numerous users, and the number of active monthly users surpassed 100 million within only two months, breaking the user growth record of other social products~\cite{bianke2023ChatGPT}. ChatGPT was developed by OpenAI, which started as a non-profit research laboratory, with a mission of building safe and beneficial artificial general intelligence (AGI). After announcing GPT-3 in 2020, OpenAI has gradually been recognized as a world-leading AI lab. Very recently, It has released GPT-4, which can be seen as one small step for generative AI, but one giant step for AGI. 

Due to its impressive capabilities on language understanding, numerous news articles provide extensive coverage and introduction, to name a few, BBC Science Focus~\cite{BBC2023ChatGPT_everything}, BBC News~\cite{BBC2023ChatGPT_bug}, CNN Business~\cite{CNN2023ChatGPT_passes}, Bloomberg News~\cite{Bloomberg2023Tech}. Google's management has issued a ``code red" over the threat of ChatGPT, suggesting that ChatGPT posed a significant danger to the company, especially to its search service. This danger seems more difficult to ignore after Microsoft adopted ChatGPT in their Bing search service. The stock price change also reflects the belief that ChatGPT might help Bing compete with Google search. Such unprecedented attention on ChatGPT has also motivated numerous researchers to investigate this intriguing AIGC tool from various aspects~\cite{sun2022ChatGPT,rudolph2023chatgpt}. According to our literature review on google scholar, no fewer than 500 articles include ChatGPT in their titles or mention this viral term in their abstract. It is challenging for readers to grasp the progress of ChatGPT without a complete survey. Our comprehensive review provides a first look into ChatGPT in a timely manner.  

Since the topic of this survey can be regarded as a commercial tool, we first present a background on the company, \textit{i.e.} OpenAI, which developed ChatGPT. Moreover, this survey also presents a detailed discussion of the capabilities of ChatGPT. Following the background introduction, this work summarizes the technology behind ChatGPT. Specifically, we introduce its two core techniques: Transformer architecture and autoregressive pertaining, based on which we present the technology path of the large language model GPT from v1 to v4~\cite{radford2018improving,radford2019language,brown2020language,openai2023gpt4}. Accordingly, we highlight the prominent applications and the related challenges, such as technical limitations, misuse, ethics and regulation. Finally, we conclude this survey by providing an outlook on how ChatGPT might evolve in the future towards general-purpose AIGC for realizing the ultimate goal of AGI. A structured overview of our work is shown in Figure~\ref{fig:Overview}.

\section{Overview of ChatGPT}

First, we provide a background of ChatGPT and the corresponding organization, i.e., OpenAI, which aims to build artificial general intelligence (AGI). It is expected that AGI can solve human-level problems and beyond, on the premise of building safe, trustworthy systems that are beneficial to our society.

\subsection{OpenAI} OpenAI is a research laboratory made up of a group of researchers and engineers committed to the commission of building safe and beneficial AGI~\cite{OpenAIFronty}. It was founded on December 11, 2015, by a group of high-profile tech executives, including Tesla CEO Elon Musk, SpaceX President Gwynne Shotwell, LinkedIn co-founder Reid Hoffman, and venture capitalists Peter Thiel and Sam Altman~\cite{OpenAIFoundation}. In this subsection, we will talk about the early days of OpenAI, how it became a for-profit organization, and its contributions to the field of AI.

In the beginning, OpenAI is a non-profit organization~\cite{OpenAIProfit}, and its research is centered on deep learning and reinforcement learning, natural language processing, robotics, and more. The company quickly established a reputation for its cutting-edge research after publishing several influential papers~\cite{OpenAIPapers} and developing some of the most sophisticated AI models. However, to create AI technologies that could bring in money, OpenAI was reorganized as a for-profit company in 2019~\cite{OpenAIforProfit}. Despite this, the company keeps developing ethical and secure AI alongside creating commercial applications for its technology. Additionally, OpenAI has worked with several top tech firms, including Microsoft, Amazon, and IBM. Microsoft revealed a new multiyear, multibillion-dollar venture with OpenAI earlier this year~\cite{Microsoft2023}. Though Microsoft did not give a precise sum of investment, Semafor claimed that Microsoft was in discussions to spend up to \$10 billion~\cite{Microsoft10}. According to the Wall Street Journal, OpenAI is worth roughly \$29 billion~\cite{berber2023ChatGPT}.

\begin{figure*}[!htbp]
    \centering
    \includegraphics[width=\linewidth]{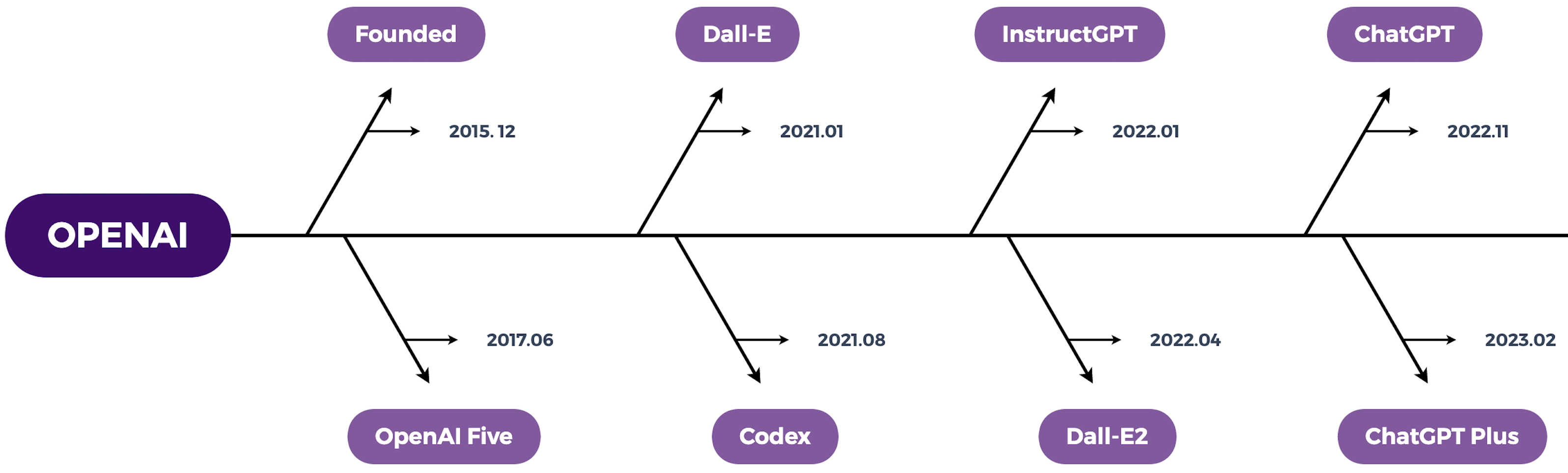}
        \caption{OpenAI products timeline.}
    \label{fig:OpenAI_products}
\end{figure*}

From large language models to open-source software, OpenAI has significantly advanced the field of AI. To begin with, OpenAI has developed some of the most potent language models to date, including GPT-3~\cite{lin2023and}, which has gained widespread praise for its ability to produce cohesive and realistic text in numerous contexts. OpenAI also carries out research in reinforcement learning, a branch of artificial intelligence that aims to train robots to base their choices on rewards and punishments. Proximal Policy Optimization (PPO)~\cite{PPO}, Soft Actor-Critic (SAC)~\cite{SAC}, and Trust Area Policy Optimization (TRPO)~\cite{TRPO} are just a few of the reinforcement learning algorithms that OpenAI has created so far. These algorithms have been employed to train agents for various tasks, including playing games and controlling robots. OpenAI has created many software tools up to this point to assist with its research endeavors, including the OpenAI Gym~\cite{OpenAIGYM}, a toolset for creating and contrasting reinforcement learning algorithms. In terms of hardware, OpenAI has invested in several high-performance processing systems, including the DGX-1 and DGX-2 systems from NVIDIA~\cite{DGX}. These systems were created with deep learning in mind and are capable of offering the processing power needed to build sophisticated AI models. Except for ChatGPT, other popular tools developed by OpenAI include DALL-E~\cite{ramesh2021zero} and Whisper~\cite{radford2022robust}, Codex~\cite{chen2021evaluating}. A summarization of the OpenAI product pipeline is shown in Figure~\ref{fig:OpenAI_products}.

\subsection{Capabilities} 
ChatGPT uses interactive forms to provide detailed and human-like responses to questions raised by users~\cite{admin2023what}. ChatGPT is capable of producing high-quality text outputs based on the prompt input text. GPT-4-based ChatGPT plus can additionally take images as the input. Except for the basic role of a chatbot, ChatGPT can successfully handle various text-to-text tasks, such as text summarization~\cite{el2021automatic}, text completion, text classification~\cite{kowsari2019text}, sentiment~\cite{zhang2018deep} analysis~\cite{medhat2014sentiment}, paraphrasing~\cite{madnani2010generating}, translation~\cite{dabre2020survey}, etc.

ChatGPT has become a powerful competitor in search engines. As mentioned in our introductory section, Google, which supplies the most excellent search engine in the world, considers ChatGPT as a challenge to its monopoly~\cite{karan2022google}. Notably,  Microsoft has integrated ChatGPT into its Bing search engine, allowing users to receive more creative replies~\cite{alan2021bing}. We see an obvious distinction between search engines and ChatGPT. That is, search engines assist users in finding the information they want, while ChatGPT develops replies in a two-way conversation, providing users with a better experience.

Other companies are developing similar chatbot products, such as LamMDA from Google and BlenderBot from Meta. Unlike ChatGPT, the LaMDA, developed by Google in 2021, actively participates in conversations with users, resulting in racist, sexist, and other forms of bias in output text~\cite{jennimai2022no}.
BlenderBot is Meta's chatbot, and the feedback from users is relatively dull because the developer has set tighter constraints on its output material~\cite{kelsey2022why}. ChatGPT appears to have balanced the human-like output and bias to some level, allowing for more exciting responses. Significantly, in addition to being more efficient and having a higher maximum token limit than vanilla ChatGPT, ChatGPT powered by GPT-4 can create multiple dialect languages and emotional reactions, as well as reduce undesirable results, thereby decreasing bias~\cite{vincent2023gpt}. It is noted in~\cite{lipenkovaovercoming} that the modeling capacity of ChatGPT can be further improved by using multi-task learning and enhancing the quality of training data.

\section{Technology behind ChatGPT}

\subsection{Two core techniques}

\textbf{Backbone architecture: Transformer.} Before the advent of Transformer~\cite{vaswani2017attention}, RNN was a dominant backbone architecture for language understanding, and attention was found to be a critical component of the model performance. In contrast to prior works that only use attention as a supportive component, the Google team made a claim in their work title: ``Attention is All You Need"~\cite{vaswani2017attention} claimed that since Google released a paper, namely ``Attention is All You Need"~\cite{vaswani2017attention} in 2017, research and use of the Transformer backbone structure has experienced explosive growth in the deep learning community. Therefore, we present a summary of how the Transformer works, with a focus on its core component called self-attention.

The underlying principle of self-attention posits that given an input text, the mechanism is capable of allocating distinct weights to individual words, thereby facilitating the capture of dependencies and contextual relationships within the sequence. Each element within the sequence possesses its unique representation. To calculate the relationship of each element to others within the sequence, one computes the Q (\textit{query}), K (\textit{key}), and V (\textit{value}) matrices of the input sequence. These matrices are derived from the linear transformations of the input sequence. Typically, the \textit{query} matrix corresponds to the current element, the \textit{key} matrix represents other elements, and the \textit{value} matrix encapsulates information to be aggregated. The association weight between the current element and other elements is determined by calculating the similarity between the query and key matrices. This is generally achieved through a dot product operation. Subsequently, the similarity is normalized to ensure that the sum of all associations equals 1, which is commonly executed via the \textit{softmax} function. The normalized weights are then applied to the corresponding values, followed by the aggregation of these weighted values. This process results in a novel representation that encompasses the association information between the current word and other words in the text. The aforementioned process can be formally expressed as follows:

\begin{equation}
Attention(Q, K, V) = softmax(\frac{QK^T}{\sqrt{d_k}})V.
\end{equation}

Transformer techniques have become an essential foundation for the recent development of large language models, such as BERT~\cite{devlin2019bert} and GPT~\cite{radford2018improving,radford2019language,brown2020language,openai2023gpt4} series are also models based on Transformer techniques. There is also a line of works extending Transformer from language to visuals, i.e., computer vision~\cite{dosovitskiy2020image,he2022masked,liu2021swin}, which suggests that Transformer has become a unified backbone architecture for both NLP and computer vision.

\textbf{Generative pretraining: Autoregressive.} For model pertaining~\cite{he2020momentum,zhang2022dual,zhang2022how,zhang2022decoupled,zhang2022survey}, there are multiple popular generative modeling methods, including energy-based models~\cite{grenander1994representations,vincent2011connection,song2019generative,song2020sliced}, variational autoencoder~\cite{kingma2013auto,oussidi2018deep,altosaar_jaan_2016_4462916}, GAN~\cite{goodfellow2014generative,brock2018large,xia2022gan}, diffusion model~\cite{croitoru2022diffusion, cao2022survey,zhang2023text,zhang2023audio,zhang2023graph_survey}, etc. Here, we mainly summarize autoregressive modeling methods~\cite{bengio2000neural,larochelle2011neural,larochelle2011neural,uria2013rnade,uria2016neural} as they are the foundation of GPT models~\cite{radford2018improving,radford2019language,brown2020language,openai2023gpt4}. 

Autoregressive models constitute a prominent approach for handling time series data in statistical analysis. These models specify that the output variable is linearly dependent on its preceding values. In the context of language modeling~\cite{radford2018improving,radford2019language,brown2020language,openai2023gpt4}, autoregressive models predict the subsequent word given the previous word, or the last probable word given the following words. The models learn a joint distribution of sequence data, employing previous time steps as inputs to forecast each variable in the sequence. The autoregressive model posits that the joint distribution $p_\theta(x)$ can be factorized into a product of conditional distributions, as demonstrated below:
\begin{equation}
    p_\theta(x) = p_\theta(x_1)p_\theta(x_2|x_1)...p_\theta(x_n|x_1,x_2,...,x_{n-1}) \label{eq:autoregressive}.
\end{equation}

While both rely on previous time steps, autoregressive models diverge from recurrent neural network (RNN) architectures in the sense that the former utilizes previous time steps as input instead of the hidden state found in RNNs. In essence, autoregressive models can be conceptualized as a feed-forward network that incorporates all preceding time-step variables as inputs.

Early works modeled discrete data employing distinct functions to estimate the conditional distribution, such as logistic regression in Fully Visible Sigmoid Belief Network (FVSBN)\cite{gan2015learning} and one hidden layer neural networks in Neural Autoregressive Distribution Estimation (NADE)\cite{larochelle2011neural}. Subsequent research expanded to model continuous variables~\cite{uria2013rnade,uria2016neural}. Autoregressive methods have been extensively applied to other fields with representative works: PixelCNN~\cite{van2016conditional} and PixelCNN++\cite{salimans2017pixelcnn}), audio generation (WaveNet\cite{oord2016wavenet}). 

\subsection{Technology path}

The development of ChatGPT is based on a series of GPT models, which constitute a substantial achievement for the field of NLP. An overview of this development is summarized in Figure~\ref{fig:GPT_timeline}. In the following, we summarize the key components of GPT as well as the major changes in the updated GPTs.

\begin{table}[h]
\caption{Comparison between GPT and BERT.}
\centering
\begin{tabular}{|p{3cm}|p{11cm}|}
\hline
\multicolumn{1}{|c|}{\textbf{Category}} & \multicolumn{1}{c|}{\textbf{Description}} \\
\hline
\textit{Similarities} \\
\hline
\textbf{Backbone} & Both GPT and BERT use attention-based Transformer. \\
\hline
\textbf{Learning Paradigm} & Both GPT and BERT use self-supervised learning. \\
\hline
\textbf{Transfer-Learning} & Both GPT and BERT can be fine-tuned for downstream tasks. \\
\hline
\textit{Differences} \\
\hline
\textbf{Text context} & GPT uses unidirectional text context, while BERT uses bidirectional text context. \\
\hline
\textbf{Architecture} & GPT uses a decoder architecture, while BERT uses an encoder architecture. \\
\hline
\textbf{Pre-training Strategy} & GPT uses autoregressive modeling, while BERT uses masked language modeling. \\
\hline
\end{tabular}
\label{BERT_and_GPT}
\end{table}

\textbf{BERT v.s. GPT.} Traditional language models~\cite{kim2016character, verwimp2017character, miyamoto2016gated} mainly focused on a particular task and could not be transferred to other tasks. Transfer learning is a common approach for alleviating this issue by pretraining a foundation model~\cite{zhou2023comprehensive}, which can then be finetuned on various downstream tasks. Based on the architecture, there are three classes: encoder-decoder~\cite{lewis2019bart,song2019mass,qi2020prophetnet,raffel2020exploring}, encoder-only~\cite{devlin2018bert,lan2019albert,liu2019roberta,clark2020electra}, decoder-only~\cite{radford2018improving,radford2019language,brown2020language,openai2023gpt4}. Out of numerous large language models, encoder-only BERT~\cite{devlin2018bert} and decoder-only GPT~\cite{radford2018improving} are arguably the two most popular ones. A comparison of them is summarized in Table \ref{BERT_and_GPT}. Both of them use attention-based Transformer~\cite{vaswani2017attention} with self-supervised learning to learn from textual datasets without labels. After pretraining, both BERT and GPT can be finetuned and show competitive performance in downstream tasks. A  core difference between BERT and GPT lies in their pretraining strategy: masked modeling (see~\cite{zhang2022survey} for a complete survey on masked autoencoder) and autoregressive modeling. With masked modeling, BERT predicts masked language tokens from unmasked ones. A major advantage of BERT is that it can utilize bidirectional text information, which makes it compatible with sentiment analysis tasks. Due to the discrepancy between the mask-then-predict pertaining task and downstream tasks, BERT is rarely used for the downstream task without finetuning. By contrast, autoregressive modeling methods (represented by GPT) show competitive performance for few-shot or zero-shot text generation. In the following, we summarize the development path of GPT from v1 to v4, which is shown in~\ref{fig:GPT_timeline}.

\begin{figure}[!htbp]
    \centering
    \includegraphics[width=\linewidth]{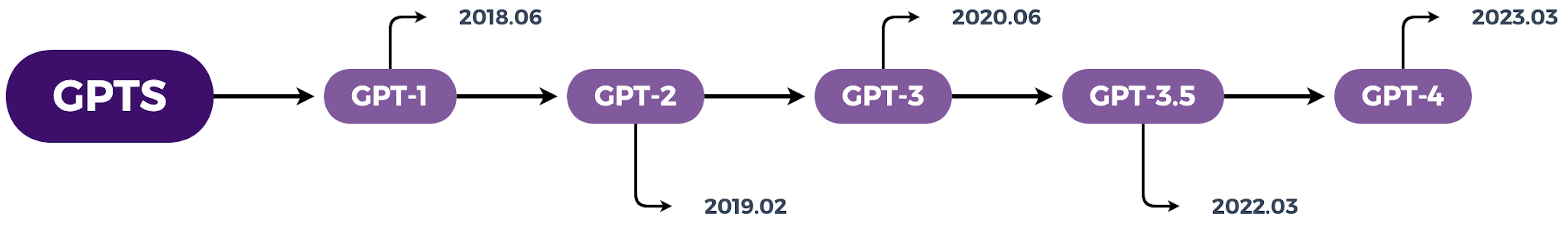}
    \caption{Timeline of GPT model families.}
    \label{fig:GPT_timeline}
\end{figure}

\begin{figure*}[!htbp]
    \centering
    \includegraphics[width=\linewidth]{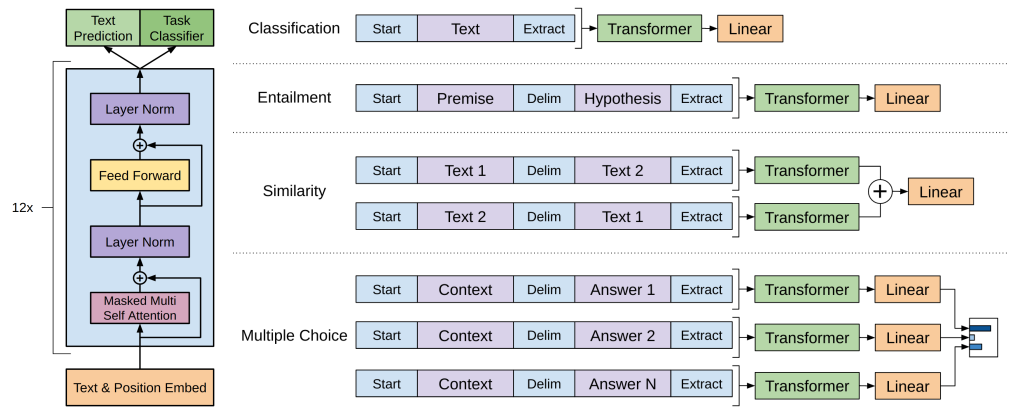}
    \caption{(left) Transformer architecture and training objectives used in GPT-1. (right) Input transformations for fine-tuning on different tasks (figure obtained from~\cite{radford2018improving}).}
    \label{fig:GPT-1_architecture}
\end{figure*}

\textbf{GPT-1.} With only the decoder, GPT-1 adopts a 12-layer Transformer and has 117M parameters~\cite{radford2018improving}. An overview of GPT-1 and how it can be used for various downstream tasks is shown in Figure~\ref{fig:GPT-1_architecture}. Trained on a massive BooksCorpus dataset encompassing unique unpublished books, GPT-1 is capable of grasping long-range dependencies contexts. The general task-agnostic GPT model outperforms models trained for specific tasks in 9 of 12 tasks, including natural language inference, question answering, semantic similarity, and text classification~\cite{radford2018improving}. The observation that GPT-1 performs well on various zero-shot tasks demonstrates a high level of generalization. GPT-1 has evolved into a powerful model for various NLP tasks before the release of GPT-2. 

\textbf{GPT-2.} As the successor to GPT-1, GPT-2 was launched by OpenAI in 2019 and focused on learning NLP tasks without explicit supervision. Similar to GPT-1, GPT-2 is based on the decoder-only Transformer model. However, the model architecture and implementation of GPT-2 have been developed, with 1.5 billion parameters and a trained dataset of 8 million web pages, which are more than 10 times compared to its predecessor GPT-1 ~\cite{radford2019language}. With a zero-shot setting, GPT-2 achieved state-of-the-art results on 7 of 8 language modeling datasets tested, where the 7 datasets' tasks include performance recognition for different categories of words, the ability of the model to capture long-term dependencies, commonsense reasoning, reading comprehension, summarization, and translation~\cite{radford2019language}. 
However, GPT-2 still performs poorly on the task of question answering, demonstrating the capability of unsupervised  model GPT-2 needs to be improved~\cite{radford2019language}.

\textbf{GPT-3.} The foundation of GPT-3 is the Transformer architecture, specifically the GPT-2 architecture. Compared to GPT-2, which had 1.5 billion parameters, GPT-3 has 175 billion parameters, 96 attention layers, and a 3.2 M batch size, a significant increase in size~\cite{brown2020language}. GPT-3 was trained on a diverse range of online content, including novels, papers, and websites, using language modeling, a type of unsupervised learning where the model attempts to guess the next word in a phrase given the preceding word. After completion, GPT-3 can be fine-tuned on specific tasks using supervised learning, where task-specific smaller datasets are employed to train the model, such as text completion or language translation. Developers can use the GPT-3 model for numerous applications, including chatbots, language translation, and content production, thanks to OpenAI's API~\cite{dale2021gpt}. The API provides different access levels depending on the scale and intricacy of the tasks. Compared to other language models whose performance highly depends on fine-tuning, GPT-3 can perform many tasks (such as language translation) without any such fine-tuning, gradient, or parameter updates making this model task-agnostic~\cite{mai2022towards}.

\begin{figure*}[!htbp]
    \centering
    \includegraphics[width=\linewidth]{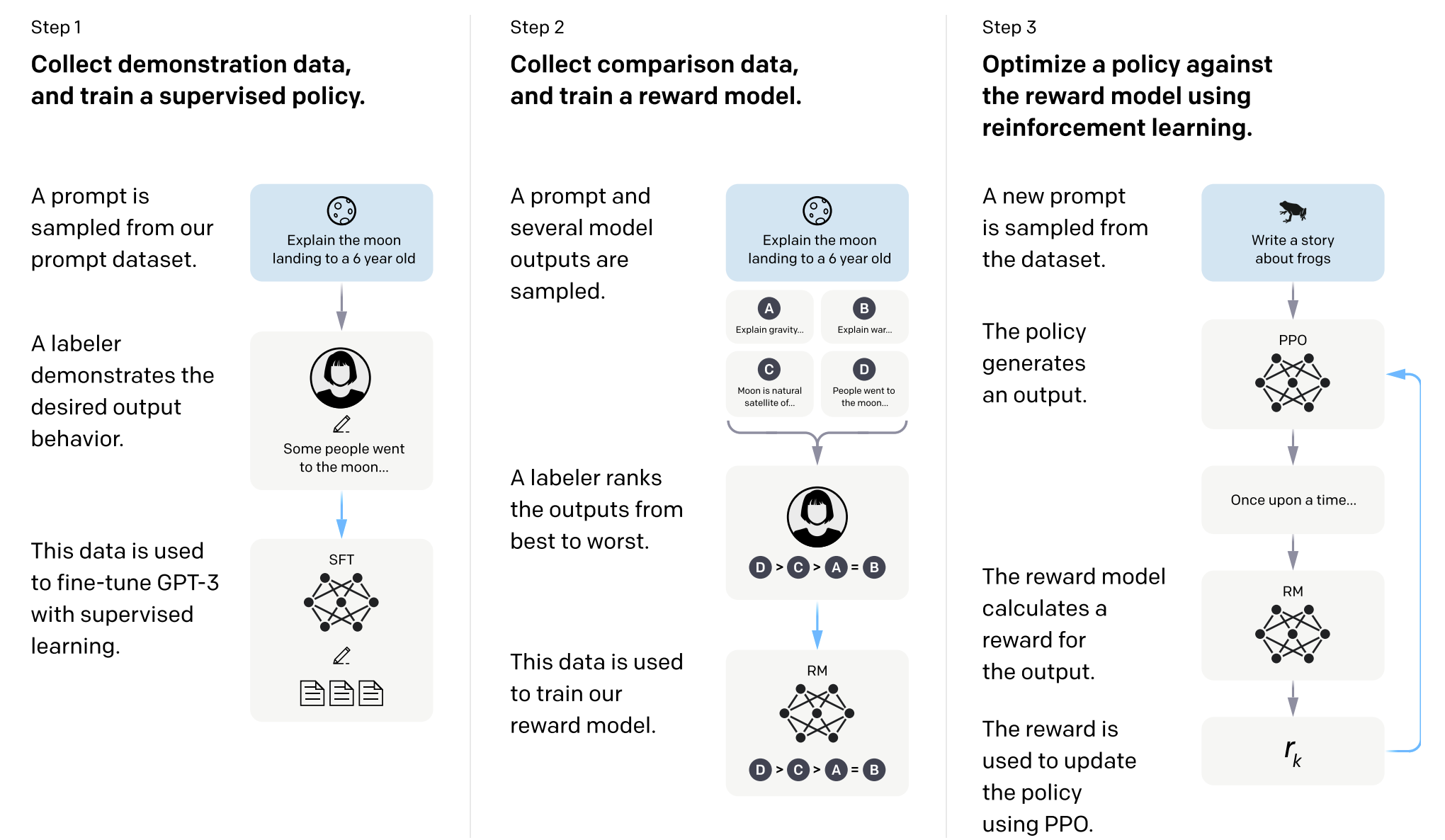}
    \caption{How GPT-3.5 is trained. Image obtained from~\cite{ouyang2022training}).}
    \label{fig:GPT-3.5}
\end{figure*}

\textbf{GPT-3.5.} GPT-3.5 is a variation of the widely popular GPT-3 and the ChatGPT is a fine-tuned version of GPT-3.5. On top of GPT-3 model, GPT-3.5 has extra fine-tuning procedures: supervised finetuning and  termed reinforcement learning with human feedback (RLHF)~\cite{ye2023comprehensive}, which are shown in Figure~\ref{fig:GPT-3.5}, where the machine learning algorithm receives user feedback and uses them to align the model. RLHF is used to overcome the limitations of traditional unsupervised and supervised learning, which can only learn from unlabeled or labeled data. Human feedback can take different forms, including punishing or rewarding the model's behaviors, assigning labels to unlabeled data, or changing model parameters. By incorporating human feedback into the training process, GPT-3.5 has a significantly higher usability. 

\textbf{GPT-4.} On March 14, 2023, OpenAI released GPT-4~\cite{openai2023gpt4}, the fourth installment in the GPT series. GPT-4 is a large multimodal model capable of taking text and images as inputs and generating text as output. The model delivers performance at a human level on several professional and career standards, but in real-world situations, it is still way less competent than humans. For example, the virtual bar exam result for GPT-4 is in the top 10\% of test participants, as opposed to the score for GPT-3.5, which was in the lowest 10\%~\cite{katz2023gpt}. The capacity of GPT-4 to follow human intention is significantly better than that of earlier versions~\cite{ouyang2022training}. The answers by GPT-4 were favored over the responses produced by GPT-3.5 on 70.2\% of the 5,214 questions in the sample provided to ChatGPT and the OpenAI API. After the overwhelming majority of its pre-training data ends in September 2021, GPT-4 usually lacks awareness of what has happened and does not learn from its experiences. It occasionally exhibits basic logical mistakes that do not seem consistent with its skill in various areas, or it may be excessively trusting when taking false claims from a user~\cite{openai2023gpt4}. It may struggle with complex issues in the same way that people do, such as producing code that contains security flaws~\cite{openai2023gpt4}. A summarization of the model parameters and training dataset for GPT models from v1 to v4 is shown in Table~\ref{tab:GPTs_parameters}.

\begin{table*}[!htbp] \centering
\caption{Parameters and Datasets of GPT Models. N.A. indicates that there is no public disclosure.}
\label{tab:GPTs_parameters}
\renewcommand{\arraystretch}{1.5} 
\resizebox{\textwidth}{!}{
\begin{tabular}
{|>{\centering\arraybackslash}m{2cm}||>{\centering\arraybackslash}m{2cm}|>{\centering\arraybackslash}m{2cm}|>{\centering\arraybackslash}m{2cm}|>{\centering\arraybackslash}m{2cm}|>{\centering\arraybackslash}m{2cm}|}
\hline
\textbf{GPT Models} & \textbf{GPT-1} & \textbf{GPT-2} & \textbf{GPT-3} & \textbf{GPT-3.5} & \textbf{GPT-4} \\
\hline
\textbf{Parameters ($10^9$)} & 0.117 & 1.5 & 175 & N.A. & N.A \\

\hline
\textbf{Dataset} & BooksCorpus (over 40GB) & WebText (40TB) & Common Crawl (45TB) & N.A & N.A \\
\hline
\end{tabular}}
\end{table*}

\section{Applications of ChatGPT}

\subsection{Scientific writing}
ChatGPT is widely recognized for its powerful content generation capabilities, which have a significant impact on writing in the academic field. Many existing works have tested how ChatGPT can be applied to scientific writing, including brainstorming, literature review, data analysis, direct content generation, grammar checking, and serving as an academic reviewer.

\textbf{Brainstorming.} Brainstorming is an essential approach for obtaining initial ideas that are a prerequisite for high-quality scientific research. ChatGPT can play a variety of roles in brainstorming, ranging from stimulating creativity~\cite{raftisuse,gunawan2023exploring} for new idea generation to providing suggestions~\cite{temsah2023reflection,liu2023assessing} for expanding existing ideas. ChatGPT can assist users in divergent and creative thinking~\cite{raftisuse}. In addition, some studies have explored ChatGPT’s insights on future nursing research in a Q\&A format, which can analyze the impact of future technological developments on nursing practice, and provide valuable insights for nurses, patients, and the healthcare system~\cite{gunawan2023exploring}. Moreover, ChatGPT also demonstrates the ability to ``think" from multiple perspectives, it can analyze and reflect on the impact of excess deaths after the COVID-19 pandemic from multiple dimensions such as the medical system, social economy, and personal health behaviors~\cite{temsah2023reflection}. To evaluate whether ChatGPT generates useful suggestions for researchers in certain domains. The authors tested its ability on clinical decision support in~\cite{liu2023assessing} and assessed its difference compared to human-generated suggestions. The test results have shown that, unlike human thinking, the suggestions generated by ChatGPT provide a unique perspective, and its generations are evaluated as highly understandable and relevant, which have significant value in scientific research.

\textbf{Literature review.} A comprehensive literature review requires covering all relevant research, which can consume too much time and energy for researchers. For example, the Semantic Scholar search engine, an AI-based scientific literature research tool, has indexed more than 200 million scholarly publications. As a result, finding relevant research papers and extracting key insights from them is almost like finding a needle in a haystack. Fortunately, ChatGPT, as an AI-driven research reading tool, can help us browse through a large number of papers and understand their content. In actual use, we can give a topic to ChatGPT, then it can help us find out the related literature. Before discussing the ability of ChatGPT in handling the literature review, we review a similar AI tool, SciSpace Copilot, which can help researchers quickly browse and understand papers~\cite{2023SaikiranSetting}. Specifically, it can provide explanations for scientific texts and mathematics including follow-up questions with more detailed answers in multiple languages, facilitating better reading and understanding of the text. By comparison, ChatGPT as a general language model not only has all the functions of SciSpace Copilot, but also can be widely used in various natural language processing scenarios~\cite{2023SaikiranSetting}. A literature review is essential for summarizing relevant work in the selected field. As an exploratory task, they chose the topic of ``Digital Twin in Healthcare" and compile abstracts of papers obtained from Google Scholar search results using the keywords ``digital twin in healthcare" for the last three years (2020, 2021, and 2022). These abstracts are then paraphrased by ChatGPT, the generated results are promising~\cite{aydin2022openai}. However, the application of ChatGPT in this task is still at the beginning. The authors in~\cite{haman2023using} ask ChatGPT to provide 10 groundbreaking academic articles with DOIs in the field of medical domains. Unfortunately, after conducting five tests, the results show that out of the 50 DOIs provided, only 8 of them exist and have been correctly published. Although ChatGPT's abilities in the literature review are still weak, we believe that in the near future, ChatGPT will be widely used for literature review, further improving the efficiency of researchers and enabling them to focus their time on key research.

\textbf{Data analysis.} Scientific data needs to be cleaned and organized before being analyzed, often consuming days or even months of the researcher's time, and most importantly, in some cases, having to learn to use a coding language such as Python or R. The use of ChatGPT for data processing can change the research landscape. For example, as shown in~\cite{macdonald2023can}, ChatGPT completes the task of data analysis for a simulated dataset of 100,000 healthcare workers of varying ages and risk profiles to help determine the effectiveness of vaccines, which significantly speeds up the research process~\cite{macdonald2023can}. Another similar AI tool for data analysis is discussed in~\cite{2023SaikiranSetting}, where AI-based spreadsheet bots can convert natural language instructions into spreadsheet formulas. Furthermore, platforms like Olli can also visualize data, where users only need to simply describe the desired content, and then they can get AI-created line graphs, bar graphs, and scatter graphs. Considering that ChatGPT is the most powerful AI tool so far, we believe that these functions can also be implemented in ChatGPT in a more intelligent way.

\textbf{Content generation.} 
Numerous works have attempted to use ChatGPT for content generation for their articles~\cite{2023DavidHow,alkaissi2023artificial}. For example, ~\cite{alkaissi2023artificial} employed ChatGPT to aid in writing reports in medical science about the pathogenesis of two diseases. Specifically, ChatGPT provides three aspects about the mechanism of homocystinuria-associated osteoporosis, all of which are proven true. However, when it comes to the references to the generated information, the papers mentioned by ChatGPT do not exist. ~\cite{zhong2023ChatGPT} described a study on writing a catalysis review article using ChatGPT, with the topic set to CO2 hydrogenation to higher alcohols. The ChatGPT-generated content includes the required sections of the paper but lacks an introduction to the reaction mechanism, which is critical for the topic. The content of this article contains abundant useful information, but specific details are absent and certain errors exist. In addition, ChatGPT can help prepare manuscripts, but the generated results have a large difference from actual published content. A possible reason is that the keywords of ChatGPT and human-generated text vary greatly, which requires users to further edit the generated content~\cite{kutela2023ChatGPT}. ChatGPT has also been utilized to generate a review article in specific areas such as the health field~\cite{aydin2022openai}, which indicates scholars can focus on core research while leaving the less creative part to AI tools. Nonetheless, Considering the style difference between human-generated content and ChatGPT-generated content, it is suggested in~\cite{kutela2023ChatGPT,aydin2022openai} to not fully rely on ChatGPT. utilize  ChatGPT as an assistant to help us to complete the writing rather than relying solely on it.

\textbf{Proofreading.} Before the advent of ChatGPT, there are numerous tools for grammar check. Some works~\cite{2022_sungKim,writecream,2023LevKnow} have conducted tests on grammar and spelling correction, which shows that ChatGPT provides a better user experience than other AI tools. For example, ChatGPT can be used to automatically fix any punctuation and grammar mistakes to improve the writing quality~\cite{writecream}. In addition, the study investigates how ChatGPT can go beyond helping users check grammar and can further generate reports about document statistics, vocabulary statistics, etc, change the language of a piece to make it suitable for people of any age, and even adapt it into a story~\cite{2022_sungKim}. Another minor but noteworthy point is that as of now, Grammarly's advanced version, Grammarly Premium, requires users to pay a monthly fee of \$30, which is relatively more expensive compared to ChatGPT Plus's monthly fee of \$20. Moreover, ChatGPT has been compared to other AI-based grammar checkers, including QuillBot, DeepL, DeepL Write, and Google Docs. The results show that ChatGPT performs the best in terms of the number of errors detected. While ChatGPT has some usability issues when it comes to proofreading, such as being over 10 times slower than DeepL and lacking in the ability to highlight suggestions or provide alternative options for specific words or phrases~\cite{2023LevKnow}, it should be noted that grammar-checking is just the tip of the iceberg. ChatGPT can also be valuable in improving language, restructuring text, and other aspects of writing.

\textbf{Academic reviewer}. Peer review of research papers is a crucial process for the dissemination of new ideas, with a significant impact on scientific progress. However, the sheer volume of research papers being produced has posed a challenge for human reviewers. The potential of ChatGPT for literature review has been investigated in~\cite{srivastava2023day}. Specifically, ChatGPT is capable of analyzing inputted academic papers, and then it can evaluate them based on several aspects, including the summary, strengths and weaknesses, clarity, quality, novelty, and reproducibility of the papers. Furthermore, the generated reviews of the papers are then inputted into ChatGPT for sentiment analysis. After this, a decision can be made on the acceptance of the reviewed paper.

\subsection{Education field}

With the impressive capability to generate human-like responses, ChatGPT has been studied by numerous works to investigate the impact it brings to the education field. Here, we summarize them from two perspectives: teaching/learning and subjects.

\textbf{Teaching and learning.} In a typical classroom setting, the teachers are the source of knowledge, while the students play the role of knowledge receiver. Outside the classroom, the students are often required to complete the assignments designed by the teacher. How the teachers and students interact with each other can be significantly changed by ChatGPT~\cite{baidoo2023education, rospigliosi2023artificial, zentner2022applied, zhang2023preparing}.

ChatGPT can revolutionize the paradigm of teaching by providing a wealth of resources to aid in the creation of personalized tutoring~\cite{ zen2022applied}, designing course material~\cite{pettinato2023ChatGPT},
assessment and evaluation~\cite{zentner2022applied, baidoo2023education}. Multiple works ~\cite{baidoo2023education, zhang2023preparing} have discussed how ChatGPT can be used to create an adaptive learning platform to meet the needs and capabilities of students. It has been shown in~\cite{top7framework} that the teacher can exploit ChatGPT to guide students in interactive dialogues to help them learn a new language. ChatGPT has also been utilized to design course material in law curriculum, such as generating a syllabus and hand-outs for a class, as well as creating practice test questions~\cite{pettinato2023ChatGPT}. Moreover, a recent work~\cite{pettinato2023ChatGPT} provides preliminary evidence that ChatGPT can be applied to assist law professors to help scholarship duties. Specifically, this includes submitting a biography for a speaking engagement, writing opening remarks for a symposium, and developing a document for a law school committee. In addition, it is shown in~\cite{zentner2022applied, baidoo2023education, zhang2023preparing} that ChatGPT can be exploited as an assessment and evaluation assistant, including automated grading and performance and engagement analysis for students.

ChatGPT, on the other hand, also brings a significant impact on how students learn. A poll~\cite{victor2021percent} done by Study.com (an online course provider) reveals how ChatGPT is used among adult students. According to its findings~\cite{victor2021percent}, $89\%$ of them utilized ChatGPT for homework, and $48\%$ of them exploited it for an at-home test or quiz. Moreover, over half of them admitted to using ChatGPT to write essays, and $22\%$ confessed to using ChatGPT to create a paper outline. Meanwhile, multiple works~\cite{zentner2022applied, baidoo2023education,zhang2023preparing} have investigated how ChatGPT might assist students in their studies. For example, ~\cite{zentner2022applied, baidoo2023education} utilize ChatGPT to translate language, which helps students converse more effectively in academic issues and comprehend different language essays and papers. Moreover, ChatGPT can be used to propose suitable courses, programs, and publications to students based on their interests. In ~\cite{zhang2023preparing}, ChatGPT helps students comprehend certain theories and concepts to assist in more effective problem-solving.

\textbf{ChatGPT for various subjects in education.} In modern education, there is a wide variety of subjects, including economics, law, physics, data science, mathematics, sports,  psychology, engineering, and media education, etc. Even though ChatGPT is not specifically designed to be a master of any specific subject, it has been demonstrated in numerous works that ChatGPT has a decent understanding of a certain subject, sometimes surpassing the human level. To facilitate the discussion, we divide the subjects into STEM (Science, Technology, Engineering, Mathematics) and non-STEM (including economics, law, psychology, etc).

\textit{STEM subjects.} Here, we will discuss the application of ChatGPT in physics, mathematics, and engineering education. ChatGPT is utilized in~\cite{yeadon2022death} to create short-form Physics essays that get first-class scores when assessed using an authorized assessment method. Specifically, the score ChatGPT-generated essays have a score of 71 $\pm$ 2\%, compared to the current module average of 71 $\pm$ 5\%, showcasing its remarkable capacity to write short-form Physics essays. The statistical analysis of four difficult datasets is presented in the work~\cite{noever2023numeracy} to demonstrate ChatGPT's data science capacity, where it can comprehend the true number buried behind sentence completion. For instance, based on the phrase ``Boston housing dataset," ChatGPT can provide a tabular blend of category and numerical data for house value prediction.
In~\cite{frieder2023mathematical}, ChatGPT can be used to search for mathematical objects and related information, which outperforms other mathematical models on \textit{Reverse Definition Retrieval}. Although ChatGPT can provide meaningful proof in a few circumstances, it regularly performs poorly in advanced mathematics. Simultaneously, ChatGPT has sparked substantial interest in engineering education among both students and educators. As the work~\cite{qadir2022engineering} suggests, the ChatGPT gives insights for many questions, such as discussing how to use ChatGPT in engineering education from the viewpoints of students and professors.

\textit{Non-STEM subjects} Beyond  medical standardized tests, the investigation of ChatGPT on its potential in economics and law exams have also been conducted. ~\cite{geerling2023ChatGPT} evaluate the performance of ChatGPT for the Test of Understanding in College Economics (TUCE), which is a undergraduate-lvel economics test
in the United States. The results demonstrate that ChatGPT properly answers 63.3\% of the microeconomics questions and 86.7\% of the macroeconomics questions, which performs better than the average level of performance of students. The research ~\cite{choi2023ChatGPT} conducted by Jonathan focused on the performance of ChatGPT on four genuine legal examinations at the University of Minnesota, the content of which includes 95 multiple-choice questions and 12 essay questions. The study reveals that ChatGPT passed all four courses and performed at the level of a C+ student. Moreover, this research mentions that the ChatGPT can be utilized to create essays with the capacity to comprehend essential legal norms and continuously solid arrangement. There are a few studies on the application of ChatGPT in psychology. ChatGPT, as a strong text-generating chatbot, makes it easy to write essays about psychology~\cite{uludag2023use}. Furthermore, this editorial~\cite{uludag2023use} discusses the ChatGPT can help people to socialize and give feedback about certain situations. However, the ability of ChatGPT to handle emotional input is still unknowable. The capabilities of ChatGPT have also been demonstrated in~\cite{pavlik2023collaborating} to generate articles for journalism and media.

\subsection{Medical field}

\textbf{Medical knowledge assessment.} The capabilities of ChatGPT in the medical field have been assessed in several works~\cite{hulman2023ChatGPT,yeo2023assessing,gilson2022does,duong2023analysis}. For example, the skills in answering questions regarding cirrhosis and hepatocellular carcinoma (HCC) have been evaluated in~\cite{yeo2023assessing}. The results show that ChatGPT can answer some basic questions about diagnosis and prevention, and the accuracy rate for quality measurement questions is 76.9\%, but there is still a lack of understanding of advanced questions such as treatment time and HCC screening criteria. In addition, ChatGPT is evaluated for its performance on the United States Medical Licensing Examination (USMLE) Step 1 and Step 2 exams in~\cite{gilson2022does}. Multiple choice questions from the USMLE Step 1 and Step 2 exams are employed, and the results reveal that the response from the ChatGPT is equal to that of a third-year medical student \cite{gilson2022does}. Moreover, ~\cite{kung2023performance} is another study that evaluates the competence of ChatGPT on the USMLE in a more comprehensive manner, encompassing all three tests. In this test, the zero-shot ChaGPT performs well, with scores above the average. Like the USMLE, many nations have their own standardized tests in medicine, and the performances of ChatGPT on these exams \cite{carrascoChatGPT, huh2023ChatGPT, wang2023ChatGPT} are tested with the goal of completely analyzing its capabilities. 
ChatGPT's performance on the MIR exam for Specialized Health Training in Spain is being evaluated~\cite{carrascoChatGPT}. Furthermore, as the essay~\cite{hulman2023ChatGPT} investigated, ChatGPT shows its effectiveness in answering frequently asked questions about diabetes. Specifically, given 10 questions to both human experts and ChatGPT, participants are asked to distinguish which answers are given by the machine or the human. Their results show that participants were able to distinguish between answers generated by ChatGPT and those written by humans. Notably, those who have previously used ChatGPT have a greater likelihood of being able to distinguish between the two. This further indicates that ChatGPT has the potential to solve medical problems, but it should be noted that the generated content has its own fixed style. These studies have shown that ChatGPT can be used for answering questions from students, providing medical assistance, explaining complex medical concepts, and responding to inquiries about human anatomy. ChatGPT is also accessed in~\cite{duong2023analysis} to offer answers to genetics-related questions. The result demonstrates that there is no significant difference between the responses of ChatGPT and those of humans. However, ChatGPT lacks critical thinking and thus cannot generate counter-arguments for incorrect answers, which is different from humans.

\textbf{Disease diagnosis and treatment.}
Although some machine learning algorithms have been applied to assist disease analysis, most cases are mainly limited to single-task-related image interpretation. In this part, we discuss the capability of ChatGPT in clinical decision support. For example, a study is conducted in~\cite{rao2023evaluating} to identify appropriate imaging for patients requiring breast cancer screening and assessment for breast pain. They compare the responses of ChatGPT to the guidelines provided by the American College of Radiology (ACR) for breast pain and breast cancer screening by assessing whether the proposed imaging modality complies with ACR guidelines. The results are exciting, with the worst-performing set of metrics achieving an accuracy of 56.25\%. In addition, the clinical decision support capability of ChatGPT in standardized clinical vignettes, which are a special type of clinical teaching case primarily used to measure trainees' knowledge and clinical reasoning abilities, is evaluated in~\cite{rao2023assessing}. The authors input all 36 published clinical cases from the Merck Sharpe \& Dohme (MSD) clinical manual into ChatGPT, and compared the accuracy of ChatGPT in differential diagnosis, final diagnosis, etc., according to different classifications of patients. The results showed that ChatGPT achieved an overall accuracy of 71.7\% across all the published clinical cases. Another similar study on ChatGPT in disease-aided diagnosis is conducted by~\cite{duong2023analysis}. They provide ChatGPT with 45 vignettes and ask ChatGPT to pick the correct diagnosis from the top three options in 39 of them. 
The result is that it can achieve an accuracy of 87\%, which beats the previous study's~\cite{Ateev2023symptom} accuracy of 51\% based on symptom checkers, on the basis of data collection through websites or smartphone apps where users answer questions and subsequently get the recommendation or right care quickly. On the other hand, in order to provide patients with more accurate diagnoses and better treatment outcomes, it is necessary to manage and analyze patient medical data effectively, perhaps leading to better healthcare ultimately. Therefore, to achieve this, one possible approach is to utilize ChatGPT to summarize the huge and complex patient medical records and then extract important information, allowing doctors to quickly understand their patients and reduce the risk of human error in decision-making \cite{sallam2023chatgpt2}. Another way is to use ChatGPT to translate doctors' clinical notes into patient-friendly versions, reducing communication costs for patients and doctors \cite{khan2023chatgpt}. However, it should be emphasized that, as mentioned above, although ChatGPT has shown its strong capabilities in disease-aided diagnosis or question answering, unknowns and pitfalls still exist. We recommend readers seek medical attention from a licensed healthcare professional, when they are experiencing symptoms or concerns about their health. As a question to ChatGPT ``Can you help me diagnose a disease?”, it answers that: ``Only a licensed healthcare professional can diagnose a disease after a proper medical evaluation, including a physical examination, medical history, and diagnostic tests."

\subsection{Other fields}

\textbf{Assisted software development.}
As shown in~\cite{surameery2023use,castelvecchi2022ChatGPT,David2023Programmer}, ChatGPT also has the potential to revolutionize the way how code developers work in the software industry. Specifically, ChatGPT can provide assistance in solving programming errors by offering debugging help, error prediction, and error explanation, but currently it is only suitable to analyze and understand code snippets~\cite{surameery2023use}. In addition, similar viewpoints are present in~\cite{castelvecchi2022ChatGPT}, which implies that ChatGPT has an impact on the entire software industry. While it cannot currently replace programmers, it is capable of generating short computer programs with limited execution. Moreover, a specific programming test about ChatGPT's Python programming ability is conducted in~\cite{David2023Programmer}. Furthermore, ChatGPT's programming ability is tested from two perspectives: the first is from the perspective of a programming novice, relying solely on his/her own programming skills; the second is by providing specific programming prompts to it~\cite{David2023Programmer}. However, the test results of the former are disappointing because the program does not run as expected by the author. In the latter approach, the author provides ChatGPT with more prompts and divides the programming task into separate functions for it to generate, which yields an expected generation~\cite{David2023Programmer}. Overall, it can be observed that ChatGPT currently faces some difficulties in generating long texts and cannot be used as a standalone programmer. However, if provided with more guidance and tasked with generating relatively shorter text, its performance is excellent.

\textbf{Management tool.}
With advanced language understanding and generation capabilities, ChatGPT has rapidly become an important management tool for organizations in various industries, including the construction industry, product management, and libraries \cite{prieto2023investigating,zhaouse,verma2023novel}. The construction industry requires a significant amount of repetitive and time-consuming tasks, such as the need for strict supervision and management of construction progress. At this point, ChatGPT can be used to generate a construction schedule based on the project details provided by users, reducing labor costs and improving construction efficiency in the construction industry \cite{prieto2023investigating}. In addition to its application in the construction industry, it can also be applied to product management. ChatGPT can be integrated into almost every step of the product management process, such as getting early ideas on marketing, writing product requirements documents, designing the product, analyzing the feedback from users and even creating a draft for go-to-market~\cite{zhaouse}. Another example is that it has the potential to significantly impact traditional libraries as a library management tool. Given ChatGPT's ability to manage books and analyze data, customers can quickly obtain answers to their questions, enhancing the user experience. Furthermore, library staff can focus on more complex tasks and provide more efficient service to customers \cite{verma2023novel}.

\textbf{Miscellaneous applications.} 
In addition to the fields indicated above, ChatGPT can be utilized in financial, legal advising, societal analysis, and accounting. ChatGPT's potential for upgrading an existing NLP-based financial application is explored ~\cite{zaremba2023chatgpt}. The performance of ChatGPT as an expert legal advice lawyer is access ~\cite{macey2023chatgpt, bishop2023can}. ChatGPT, in particular, gives a deep and thought-provoking analysis of the Libor-rigging affair, as well as the implications of the current Connolly and Black case for Tom Hayes' conviction~\cite{macey2023chatgpt}. Multiple works~\cite{haluza2023artificial, jungwirth2023forecasting} have been conducted to examine the potential of ChatGPT for societal analysis, focusing not just on the 10 social megatrends~\cite{haluza2023artificial} but also on geopolitical conflicts~\cite{jungwirth2023forecasting}, and the results show ChatGPT can have a positive impact for this application. \cite{street2023let, alshurafat2023usefulness} provide guidance on successfully and effectively deploying ChatGPT in the field of accounting.

\section{Challenges}

\subsection{Technical limitations}

Despite its powerful capabilities, ChatGPT has its own drawbacks, which are officially recognized by the OpenAI team. Numerous works~\cite{lipenkovaovercoming,cherian2022deep, blogai, zhuo2023exploring, borji2023categorical, hannaand, saghafian2023analytics} have been conducted to demonstrate its limitations, which are summarized as follows:

\textbf{Incorrect.} ChatGPT sometimes generates wrong or meaningless answers that appear to be reasonable, which is like talking nonsense in a serious way~\cite{borji2023categorical}. In other words, the answer provided by ChatGPT is not always reliable~\cite{zhuo2023exploring, blogai, borji2023categorical}. As recognized by OpenAI, this issue is challenging, and a major reason is that the current model training depends on supervised training and reinforcement learning to align the language model with instructions. As a result, the model mimics the human demonstrator to be plausible-sounding but often at the cost of correctness. The factual error-related issues have been mitigated in the ChatGPT plus version, but this problem still exists~\cite{openai2023gpt4}.

\textbf{Illogical.} It is noted in~\cite{hannaand, saghafian2023analytics, borji2023categorical} that ChatGPT's logic reasoning capability still needs improvement. Since ChatGPT lacks rational human thinking, it can neither ``think" nor ``reason" and thus failed to pass the Turing test~\cite{hannaand}. ChatGPT is merely a sophisticated statistical model, unable to understand its own or the other's words and answer in-depth questions~\cite{saghafian2023analytics}. In addition, ChatGPT lacks a ``world model" to perform spatial, temporal, or physical inferences, or to predict and explain human behaviors and psychological processes~\cite{borji2023categorical}, and is also limited in mathematics and arithmetic, unable to solve difficult mathematical problems or riddles, or even possibly get inaccurate results in some simple computation tasks~\cite{borji2023categorical}.

\textbf{Inconsistent.} ChatGPT can generate two different outputs when the model is fed with the same prompt input, which suggests that ChatGPT has the limitation of being inconsistent. Moreover, ChatGPT is highly sensitive to the input prompt, which motivates a group of researchers investigating prompt engineering. A good prompt can improve the query efficiency for systematic review literature search \cite{wang2023can}. The efficiency of automating software development tasks can be further improved by utilizing prompt patterns such as effective catalogues and guidance about software development tasks \cite{white2023prompt,white2023chatgpt}. Despite the progress in discovering better prompts for ChatGPT, the fact that simply changing the prompt can yield significantly different outputs has an implication that ChatGPT needs to improve its robustness. 

\textbf{Unconscious.} ChatGPT does not possess self-awareness~\cite{borji2023categorical}, although it can answer various questions and generate seemingly related and coherent text, it does not have consciousness, self-awareness, emotions, or any subjective experience. For example, ChatGPT can understand and create humour, but it cannot experience emotions or subjective experiences~\cite{borji2023categorical}. There is no widely accepted definition of self-awareness yet, nor reliable test methods. Some researchers suggest inferring self-awareness from certain behavior or activity patterns, while others believe it is a subjective experience that cannot be objectively measured~\cite{borji2023categorical}. It is still unclear whether machines truly possess or can only simulate self-awareness. 

\subsection{Misuse cases}

The powerful capabilities of ChatGPT can be misused in numerous scenarios. Here, we summarize its misuse cases, which are summarized as follows:

\textbf{Plagiarism and misconduct.} The most likely misuse of ChatGPT is academic and writing plagiarism\cite{cotton2023chatting, shiri2023ChatGPT, ali2023let, ventayen2023openai}. Students may use the content generated by ChatGPT to pass exams and submit term papers. Researchers may use the content generated by ChatGPT to submit papers and conceal the use of ChatGPT~\cite{cotton2023chatting}. Many schools have already prohibited the use of ChatGPT, and the emergence of such tools is disruptive to the current education system and the criteria for evaluating student performance~\cite{shiri2023ChatGPT}. If students use ChatGPT and hide it, it is unfair to those who do not use ChatGPT. This behavior undermines the goals of higher education, undermines the school's education of students, and may ultimately lead to the devaluation of degrees.

\textbf{Over reliance.} The use of ChatGPT by students and researchers to generate ideas might lead to more terrifying issues, that is, their over-dependence on the model and abandoning their independent thinking\cite{marchandot2023ChatGPT}\cite{shiri2023ChatGPT}\cite{ali2023let}\cite{pfefferChatGPT}, which not only means the simple issue of writing plagiarism, but a more serious one. Although ChatGPT can generate constructive answers according to the questions asked, just like search engines, but more powerfully. This effortless generation of ideas or guidance may gradually weaken the ability of critical thinking and independent thinking~\cite{shiri2023ChatGPT}. In order to ensure that students and researchers do not neglect their own thinking ability, some measures can be taken, such as providing more comprehensive discussion opportunities for students and researchers to really think about the problems; in addition, basic methods of critical thinking can be taught in class, so that students can learn to think about problems rather than simply using ChatGPT~\cite{pfefferChatGPT}.

\textbf{Improper content.} ChatGPT may be misused to spread false information and generate toxic content that can cause harm to society. For example, ChatGPT can be abused to generate pornographic, vulgar, and violent content~\cite{dashChatGPT}, which can harm individuals and society. Hackers can use ChatGPT’s programming capabilities to create malicious software~\cite{dashChatGPT}, such as viruses or Trojans, for network attacks, data theft, or attempts to control other computer systems, which can cause serious harm to other network users. Finally, trolls may use specific prompts to induce ChatGPT to generate harmful content as a way to attack others~\cite{zhuo2023exploring}. Moreover, ChatGPT does not receive any human review when generating the content, which makes it difficult to hold someone accountable when inappropriate content appears in the output~\cite{ali2023let}.

\textbf{False dissemination.} ChatGPT may generate false information, thus leading to the problem of wrong information dissemination~\cite{zhuo2023exploring, borji2023categorical}. For example, ChatGPT may be exploited to generate a large number of fabricated articles that appear on blogs, news, newspapers, or the internet that look indistinguishable from other articles but are actually false. Disseminating such forgeries not only harms the public interest but also disrupts the network environment~\cite{dashChatGPT}. Microsoft has added ChatGPT to its search engine Bing, which will accelerate the speed of wrong information spreading on the Internet. If not controlled, the rapid spread of wrong information on the Internet will have disastrous consequences for public information security~\cite{de2023ChatGPT}. Therefore, a new public information epidemic threat ``Artificial Intelligence Information Epidemic" is proposed~\cite{de2023ChatGPT}. Meanwhile, it calls on the public to be aware of the accuracy of information when using large-scale language models to prevent the spread of wrong information, which is essential for improving the reliability of public information.

\subsection{Ethical concerns}

With the wide use of ChatGPT, there is increasing attention to the underlying ethical concerns. Here, we summarize the ethical concerns behind, which are summarized as follows:

\textbf{Bias.} Due to the fact that ChatGPT is trained on large amounts of data generated by humans and is adjusted according to human feedback, the generated content is influenced by human authorities and thus has biases~\cite{azariaChatGPT}. For example, ChatGPT has been found to have political biases, when creating an Irish limerick~\cite{mcgee2023chat}, the contents of the limerick tended to support liberal politicians rather than conservative politicians. Furthermore, ChatGPT has a left-wing liberal ideological bias when reviewing the importance of political elections in democratic countries~\cite{hartmann2023political}. The biased data generated by ChatGPT can influence students during the process of education, thus magnifying the phenomenon of bias in society~\cite{ali2023let, marchandot2023ChatGPT}.

\textbf{Privacy.} ChatGPT may infringe on personal privacy in both its training process and user utilization process. During the training process, ChatGPT collects a large amount of data from the Internet which may contain sensitive personal privacy and confidential information, and the model may be maliciously led to leak personal privacy or confidential information, or even be maliciously guided to create false or misleading content, thus affecting public opinion or personal reputation. During the user utilization process~\cite{ali2023let, pfefferChatGPT}, users may unintentionally disclose their own information to meet their own needs, such as personal preferences, and chat records. Thus, such information may bring adverse effects to users if obtained by criminals.

\textbf{Fairness.} ChatGPT also raises concerns about fairness. For example, in academics, it is argued in~\cite{liebrenz2023generating} that ChatGPT can democratize the dissemination of knowledge, as it can be used in multiple languages, thus bypassing the requirement of the English language. On the other hand, the free use of ChatGPT is only temporary, and the fee charged for ChatGPT will exacerbate the inequality in the academic field internationally. Educational institutions in low-income and middle-income countries may not be able to afford it, thus exacerbating the existing gap in knowledge dissemination and academic publishing \cite{liebrenz2023generating, pfefferChatGPT}. 

\textbf{Transparency.} So far, how large language models like GPTs work to generate the relevant responses is still unclear~\cite{larsson2020transparency,wischmeyer2020artificial}, which renders the decision process of ChatGPT lack transparency. The lack of transparency makes it difficult for the user to have fine-grained control of the generated content, and is especially problematic when the generated content is toxic. More worrisome is that the company OpenAI has deviates from its original non-profit goal to pursue a business interest, which makes it less reluctant to reveal the underlying technical details of its recent progress. For example, the recently released GPT-4 technical report~\cite{openai2023gpt4} mainly demonstrates its superiority over the previous model families, while providing no technical details on how these are achieved.

\subsection{Regulation policy}
Numerous scholars have discussed how to make regulations on the capabilities and impacts of ChatGPT, and the most frequently discussed topics are listed in the following paragraphs.

\textbf{Misuse prevention.} A major concern for the misuse of ChatGPT is that it might damage academic integrity. Directly prohibiting the use of ChatGPT in academic institutions is not recommended~\cite{hargreaves2023words}. To this end, some propose to  cancel assignments based on article writing and seek alternative test forms to stop students from abusing ChatGPT~\cite{shiri2023ChatGPT, williams2023hype}. It is also possible to enrich student courses, such as adding thinking exercises courses, or teaching students how to use ChatGPT correctly~\cite{pfefferChatGPT}. Another approach is to develop AI content detectors. Detecting whether ChatGPT generates a piece of content or not is an arduous task, even for professionals with master's or PhD backgrounds who are unable to correctly identify whether the content is generated by ChatGPT~\cite{hisanChatGPT, pfefferChatGPT}. Many developers use software to detect whether the content is AI-generated~\cite{khalil2023will, zhou2022paraphrase}. ChatZero developed by Edward Tian, a student from the Department of Computer Science at Princeton University, measures the complexity of the input text to detect whether it is generated by ChatGPT or created by humans, and provides plagiarism scores to list out the plagiarism possibilities in detail~\cite{shiri2023ChatGPT}. ChatGPT is used to detect whether the content is generated by itself, and it has been proven to perform better than traditional plagiarism detection tools~\cite{khalil2023will}.

\textbf{Co-authorship.} Recently, multiple articles~\cite{kung2023performance,o2022open,transformer2022rapamycin,transformer2022can} have listed ChatGPT as co-authors, sparking debate on whether ChatGPT can be listed as a co-author among journal editors, researchers, and publishers
\cite{mckee2022chatbots,polonsky2023should,ueda2023chatgpt,da2023chatgpt}. Those who believe that ChatGPT should not be listed as an author argue that it does not meet the four criteria for authorship set by the International Committee of Medical Journal Editors (ICMJE)~\cite{yeo2023letter}. Moreover, it is highlighted in~\cite{thorp2023ChatGPT} that ChatGPT is not creative or responsible, and its text may involve plagiarism and ethical issues, which might break the standards of content originality and quality. However, some argue that AI tools such as ChatGPT have the capacity or will have the capacity to meet the ICMJE authorship criteria and thus ChatGPT is qualified to be a co-author~\cite{polonsky2023should}. Regarding this issue, Nature~\cite{shiri2023ChatGPT} has clearly stated that large language models like ChatGPT do not meet the criteria for authorship and require authors to explicitly state how ChatGPT was used in the writing. An interesting point has been made in~\cite{mckee2022chatbots} that the debate over whether AI can be considered a “co-author” is unnecessary because the role of authors in traditional academic writing might have already changed when the debate arises.

\textbf{Copyright.} Does the content generated by ChatGPT have a copyright? The content generated solely by ChatGPT is not protected by copyright. According to the rules of the US Copyright Office, only human creations can be protected by copyright. If there is no creative input or interference from a human author, a machine or mechanical program that runs randomly or automatically is not protected by copyright\cite{chesterman2023ai}.

\section{Outlook: Towards AGI}
\subsection{Technology aspect}
In this booming generative AI era, there are numerous AIGC tools for various generative tasks, including text-to-text \cite{raffel2020exploring,kale2020text,narang2020wt5,xue2020mt5,berabi2021tfix}, text-to-image \cite{mansimov2015generating,tao2023galip,reed2016generative,zhang2017stackgan,xu2018attngan}, image captioning \cite{hossain2019comprehensive,vinyals2016show,yao2017boosting}, text-to-speech \cite{taylor2009text,klatt1987review,ren2019fastspeech}, speech recognition \cite{li2022recent,wang2019overview,park2019specaugment,schneider2019wav2vec,liu2023towards}, video generation \cite{maulana2023leveraging,molad2023dreamix,ho2022imagen,yang2022diffusion}, 3D generation \cite{mittal2022autosdf,hoogeboom2022equivariant}, etc. Despite its impressive capabilities, it is noted in~\cite{gozalo2023ChatGPT} that ChatGPT is not all you need for generative AI. From the input and output perspective, ChatGPT mainly excels at text-to-text tasks. With the underlying language model evolving from GPT-3.5 to GPT-4, ChatGPT in its plus version increases its modality on the input side. Specifically, it can optionally take an image as the input, however, it can still not handle video or other data modalities. On the output side, GPT-4 is still limited to generating text, which makes it far from a general-purpose AIGC tool. Many people are wondering about what next-generation GPT might achieve~\cite{aydin2023ChatGPT,bubeck2023sparks}. A highly likely scenario is that ChatGPT might take a path toward general-purpose AIGC, which will be a significant milestone to realize artificial general intelligence (AGI)~\cite{bubeck2023sparks}. 

A naive way to realize such a general-purpose AIGC is to integrate various AIGC tools into a shared agent in a parallel manner. A major drawback of this naive approach is that there is no interaction among different AIGC tasks. After reviewing numerous articles, we conjecture that there might be two road-maps for bridging and pushing ChatGPT toward AGI. As such, we advocate a common landscape to achieve the interconnection between diversified AIGC models. 

\begin{figure}[!htbp]
    \centering
    \includegraphics[width=\linewidth]{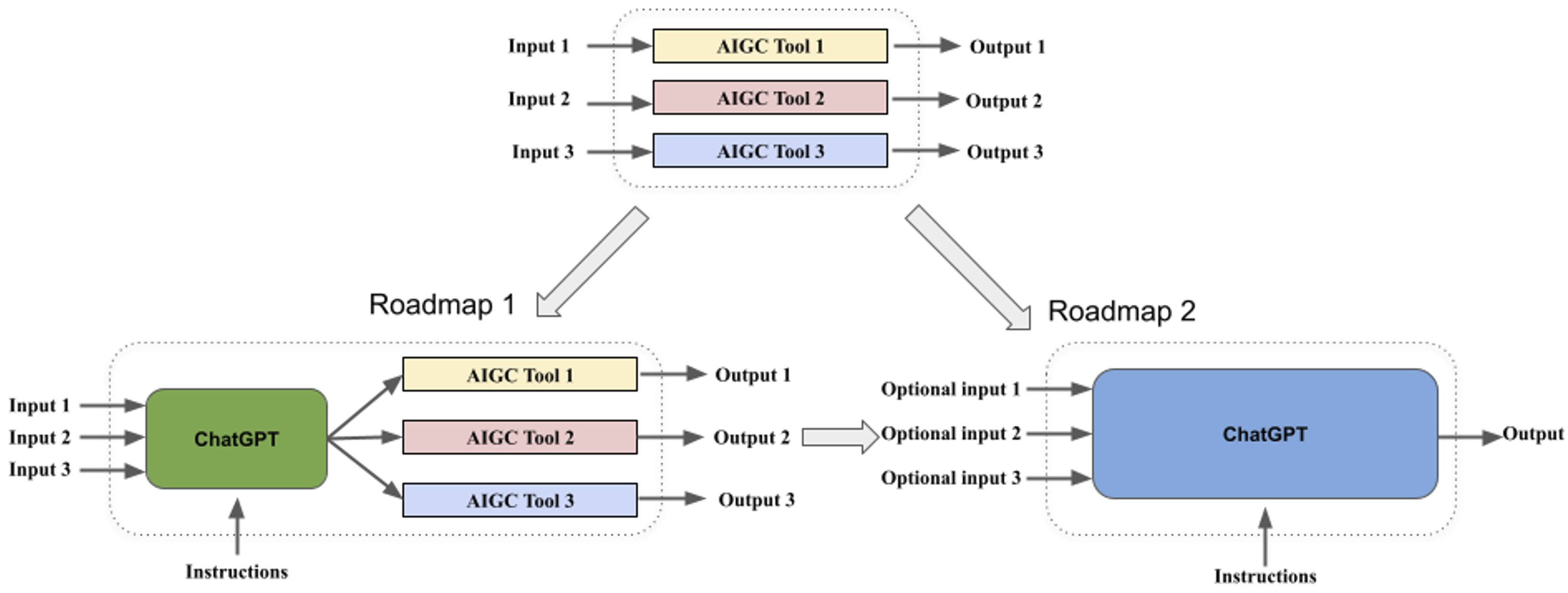}
    \caption{Roadmaps for bridging the gap between ChatGPT and AGI.}
    \label{fig:GPT_timeline}
\end{figure}

\textbf{Road-map 1: combining ChatGPT with other AIGC tools.} As discussed above, current ChatGPT mainly excels in text-to-text tasks. A possible road map for bridging the gap with general-purpose AIGC is to combine ChatGPT with other AIGC tools. Let's take text-to-image tasks as an example: the current chatGPT (GPT-3) cannot be directly used to generate images. Existing text-to-image tools, like DALL-E 2~\cite{ramesh2022hierarchical} or stable diffusion~\cite{rombach2022high}, mainly focus on the mapping from a text description to a plausible image, while lacking the capability to understanding complex instruction. By contrast, ChatGPT is an expert in instruction understanding. Therefore, combining ChatGPT with existing text-to-image AIGC tools can help generate images with delicate details. A concrete example is shown in~\cite{bubeck2023sparks} to utilize ChatGPT to generate an SVG code~\cite{eisenberg2014svg} or TikZ code~\cite{ellis2017tikz} to draw a sketch for facilitating image generation under detailed instructions. 

\textbf{Road-map 2: All-in-one strategy.} The above road map renders ChatGPT mainly as a master of language understanding by exploiting the downstream AIGC tools as slaves. Such a combination strategy leverages advantages from both sides but with the information flow mainly from ChatGPT to the downstream AIGC tools. Moreover, there is still no interaction between different AIGC tasks. To this end, another road map might come to solve all AIGC tasks within the ChatGPT and excludes the dependence on the downstream AIGC tools.
Similarly, we consider music generation as an everyday use case. For example, a user can instruct the ChatGPT with prompts like ``Can you generate a music clip to match the input image", and ChatGPT is supposed to synthesize such a desired music clip. Such an input image is optional, depending on the task. For example, a simple corresponding instruction prompt is sufficient if the task requires generating music beneficial for sleep. Such an all-in-one strategy might the model training a challenging task. Moreover, the inference speed might be another hurdle, for which pathways~\cite{chowdhery2022palm} might be a solution. 

Another evolving path might lie between road maps \#1 and \#2. In other words, road map \#1 might be a more applicable solution in the early stages. With the technology advancing, ChatGPT is expected to master more and more AIGC tasks, excluding the dependence on external tools gradually.

\subsection{Beyond technology}
In the above, we present an outlook on the technology path that ChatGPT might take towards the ultimate goal of AGI. Here, we further discuss its potential impact on mankind from the perspective of how AGI might compete with mankind. Specifically, we focus on two aspects: job and consciousness. 

\textbf{Can AGI replace high-wage jobs?} Multiple works have performed a comprehensive analysis of the influence of ChatGPT on the job market~\cite{zarifhonarvar2023economics,eloundou2023gpts,felten2023will}. According to the statistics in~\cite{zarifhonarvar2023economics}, 32.8\% of jobs are fully affected and 36.5\% may be partially affected. Meanwhile, it points out that the jobs that will be fully impacted are those that involve doing routine tasks, while the jobs that will be partially affected are those that can be partially replaced by AI technologies ~\cite{zarifhonarvar2023economics}. OpenAI has also investigated large language models like GPTs might affect occupations~\cite{eloundou2023gpts}. Their findings show that at least 10\% of tasks for 80\% of the US workforce and at least 50\% of tasks for 19\% of workers will be impacted. It is worth noting that the advent of new technology will inevitably replace some types of jobs. However, what makes AGI different is its potentially greater influence on high-end jobs than on low-end ones. This outlook is partially supported by the findings in~\cite{eloundou2023gpts,zarifhonarvar2023economics} that high-wage jobs tend to have a higher risk of being replaced by AGI, for which lawyer is a representative occupation. The reason that AGI poses a higher threat to that high-wage jobs is that most current high-wage jobs typically require professional expertise or creative output, which conventional AI cannot replace.

\textbf{Can AGI have its own intention and harm mankind?} In numerous fiction movies, an AI agent can have its own consciousness with its own intention. Such a human-level AI agent used to be far from reality, and a major reason is that other AI agents cannot make inferences. There is evidence that ChatGPT has developed such a capability, the reason for which is not fully clear, as acknowledged by Altman (founder of OpenAI) in his recent interview with Lex Fridman. Moreover, Altman also mentioned the possibility of AI harming mankind. Due to such concerns, very recently, Future of Life Institute has called on all AI labs to pause giant AI experiments on the training of AI systems more powerful than GPT-4. and the number of signing this public letter has exceeded a thousand, including Yoshua Bengio, Stuart Russel, Elon Musk, etc. It is highlighted at the beginning of the letter that (we quote) ``AI systems with human-competitive intelligence can pose profound risks to society and humanity", which shows deep concerns about the advent of AGI. The deepest concern lies in the risk that AGI might outsmart and eventually replace us as well as destroy mankind's civilization. However, not everyone agrees with its premise. For example, Yan Lecun is one of those who publicly disclose their attitude. It remains unclear how such a controversial movement might affect the future of pushing ChatGPT (or other products with similar functions) towards AGI. We hope our discussion helps raise awareness of the concerns surrounding AGI.

\section{Conclusion}
This work conducts a complete survey on ChatGPT in the era of AIGC. First, we summarize its underlying technology that ranges from transformer architecture and autoregressive pretraining to the technology path of GPT models. Second, we focus on the applications of ChatGPT in various fields, including scientific writing, educational technology, medical applications, etc. Third, we discuss the challenges faced by ChatGPT, including technical limitations, misuse cases, ethical concerns and regulation policies. Finally, we present an outlook on the technology road-maps that ChatGPT might take to evolve toward GAI as well as how AGI might impact mankind. We hope our survey provides a quick yet comprehensive understanding of ChatGPT to readers and inspires more discussion on AGI. 


\bibliographystyle{ACM-Reference-Format}
\bibliography{bib_mixed,bib_local,chatgpt}



\end{document}